\documentclass[twocolumn,showpacs,preprintnumbers,amsmath,amssymb]{revtex4}
\usepackage{graphicx}% Include figure files
\usepackage{dcolumn}% Align table columns on decimal point
\usepackage{bm}% bold math
\def\bea {\begin{eqnarray}}
\def\eea {\end{eqnarray}}

\def\be {\begin{equation}}
\def\ee {\end{equation}}

\begin{document}
%\begin{center}
%\huge{\bf Draft}
%\end{center}
\title{
The horn in the kaon to pion ratio
}

\author{Jajati  K. Nayak, Sarmistha Banik and Jan-e Alam}

\medskip

\affiliation{Variable Energy Cyclotron Centre,\\
 1/AF, Bidhan Nagar, 
Kolkata - 700064, India}.

\date{\today}

\begin{abstract}
A  microscopic approach has been employed to study the kaon productions in 
heavy ion collisions. The momentum integrated Boltzmann equation has been 
used to study the evolution of strangeness in the system formed in heavy 
ion collision at relativistic energies. The kaon productions have been 
calculated for different centre of mass energies ($\sqrt{s_{\mathrm {NN}}}$) 
ranging from AGS to RHIC. The results have been compared with available 
experimental data. We obtain a non-monotonic horn like structure for 
$K^+/\pi^+$ when plotted with $\sqrt{s_{\mathrm {NN}}}$ with the assumption 
of an initial partonic phase beyond a certain threshold  in $\sqrt{s_{NN}}$. 
However, a monotonic rise of $K^+/\pi^+$ 
is observed when a hadronic initial state is assumed for all 
$\sqrt{s_{\mathrm {NN}}}$.  Experimental values of  $K^-/\pi^-$ are also 
reproduced within the ambit of the same formalism.
Results from scenarios where the strange quarks and hadrons are formed
in equilibrium and evolves with and without secondary productions 
have also been presented. 

\end{abstract}

\maketitle

\section{Introduction}
The lattice simulation of Quantum Chromodynamic  equation of state (EoS) 
predicts that the properties of nuclear matter at extreme
densities and/or temperatures are governed by the partonic degrees of 
freedom~\cite{lattice1,lattice2,lattice3}.  
A series of experiments have been performed~\cite{npa2005} and planned
~\cite{lhc} to produce such a partonic 
state of matter, called Quark Gluon Plasma (QGP) by colliding nuclei at 
ultra-relativistic energies. 
Rigorous experimental and theoretical efforts
are on to create and detect such a novel state of matter~\cite{qm08}. 
Various signals have been proposed for the detection of QGP
 - the pros and cons of these signals
are matter of intense debate. The study of the ratio, $R^+\equiv K^+/\pi^+$
is one such currently debated issue. 
$R^+$ is measured experimentally~\cite{na49alt,star,brahms,na49afa} 
as a function of centre of mass 
energy ($\sqrt{s_{NN}}$). It is observed that the  $R^+$
increases with $\sqrt{s_{NN}}$ and then decreases beyond a certain
value of $\sqrt{s_{NN}}$ giving rise to a horn like structure, 
whereas the ratio, $R^-\equiv K^-/\pi^-$ increases faster 
at lower $\sqrt{s_{NN}}$ and tend to saturate 
at higher $\sqrt{s_{NN}}$.

Explanation of this structure has ignited intense theoretical activities
~\cite{GG,BT,sgupta,CORS,MG,andronic,jajati,LR,tawfik}. 
Several authors have attempted to reproduce the $K^+/\pi^+$ 
ratio using different approaches. 
While the authors in~\cite{BT} use a hadronic kinetic model,
in Ref.~\cite{sgupta} high mass unknown hadronic resonances 
have been introduced through Hagedorn formula to describe the data.
In Ref.~\cite{CORS} a transition from a baryon dominated
system at low energy to a meson dominated system at
higher energy  has been assumed to reproduce the ratio $K^+/\pi^+$. 
The release of color degrees of freedom is assumed in~\cite{GG}
beyond a threshold in $\sqrt{s_{\mathrm {NN}}}$ (resulting in
large pion productions) or the production of larger
number of pions than kaons from higher mass resonance decays has also
been employed~\cite{andronic}  to explain the data.  In the present work 
we employ a microscopic model for the productions and evolution of strange
quarks and hadrons depending on the collision energy.
Here we examine whether the $K^+/\pi^+$
experimental data can differentiate between the following two initial
conditions or two scenarios - after the collisions the system 
is formed in: (I) the hadronic
phase for all $\sqrt{s_{\mathrm {NN}}}$ or (II) the
partonic phase beyond a certain threshold in $\sqrt{s_{\mathrm {NN}}}$.
Other possibilities like formation  of strangeness in complete 
thermal equilibrium and evolution in space time 
(III) without and (IV) with secondary productions
of quarks and hadrons have been considered. (V) Results for  an ideal case of 
zero strangeness in the initial state has also been presented. 

In the context of strangeness enhancement as signal of QGP formation,
similar approach {\it i.e.} the assumption of an initial state
where non-strange sectors are in equilibrium but the strange 
degrees of freedom are out of equilibrium (having
density much below their equilibrium values) were considered in the 1980's. 
The strangeness
production in a deconfined (partonic) phase is enhanced compared to
the their production in the confined (hadronic) phase primarily 
because even the lightest strange hadrons, the kaons are much heavier
than the strange quarks. 
Moreover, the  strange quark has more
degrees  of freedom (six) in a deconfined matter 
compared to kaons. Therefore,  
the strangeness production during the space time evolution
of the system for partonic initial state will be enhanced
compared to the hadronic initial state, hence the enhanced
production of strangeness   
could be an efficient signal for deconfinement~\cite{rafelski,kmr,RL99}. 
In contrast to these studies 
Gazdzicki and Gorenstein ~\cite{GG,MG2,MG02}
within the ambit of statistical model 
considered the strangeness production where both 
the strange and non-strange degrees of freedom are in thermal 
equilibrium and the production of strangeness during the
expansion stage is ignored. In the present work we would
like to compare the results on  kaon to pion ratio 
from these two contrasting scenarios.

We assume that the non-strange quarks and hadrons are in complete thermal
(both kinetic and chemical) equilibrium and the strange quarks and 
strange hadrons are away from chemical equilibrium. Therefore, the
evolution of the strange sector of the system is governed by the
interactions between the equilibrium and non-equilibrium degrees of freedom.
The momentum integrated Boltzmann equation provides a possible 
framework for such studies. Similar approach has been used
to study the sequential freeze-out of elementary particles
in the early universe~\cite{kolbandturner}.

For the strangeness productions in the partonic phase 
we consider the processes of gluon fusion and light quarks annihilation.
For the production of $K^+$ and $K^-$ an exhaustive set 
of reactions involving thermal baryons and mesons have been considered. 
The time evolution of the densities are governed by the
Boltzmann equation. 
%The freeze-out parameters are constrained by the experimental data.
 
\par 
The paper is organized as follows. In the next section 
the rate of strangeness productions in the partonic and hadronic phases are
discussed. The space time evolution of the system is presented
in section III. Results are presented in section IV 
and finally section V is devoted to summary and conclusion.
%%%%%%%%%%%%%%%%%%%%%%%%%%%%%%%%%%%%%%%%%%%%%%%%%%%%%%%%%%%%%%%%%%%%%
\section{Strangeness productions}
The productions of $s$ and $\bar{s}$ in the QGP and the $K^+$ and $K^-$ in
the hadronic system are discussed below.
%%%%%%%%%%%%%%%%%%%%%%%%%%%%%%%%%%%%%%%%%%%%%%%%%%%%%%%%%%%%%%%%%%%%%
\subsection{Strange quark productions in the QGP}
%%%%%%%%%%%%%%%%%%%%%%%%%%%%%%%%%%%%%%%%%%%%%%%%%%%%%%%%%%%%%%%%%%%%%
The two main processes for the strange quark productions
are gluon fusion ($gg\rightarrow s\bar{s})$ and quark($q$)-antiquark
($\bar{q}$) annihilations ($q\bar{q}\rightarrow s\bar{s}$).
The cross sections in the lowest order QCD
is given by~\cite{rafelski}: 
\be
\sigma_{q\bar q \rightarrow s\bar s}=\frac{8\pi \alpha_s^2}{27s} 
(1+\frac{2m^2}{s}) w(s)
\ee
 and 
\be
\sigma_{gg \rightarrow s\bar s}=\frac{2\pi \alpha_s^2}{3s}
[G(s){\mathrm tanh}^{-1} w(s)-{\frac{7}{8}+\frac{31m^2}{8s}}w(s)] 
\ee
where 
$m$ is the mass of strange quark, 
$s=(p_1+p_2)^2$, is the square of the centre of mass
energy of the colliding particles,  $p_i$ 
are the four momenta of incoming particles,
$G=1+4m^2/s+m^4/s^2$,
$w(s)$=$(1-4m^2/s)$ and $\alpha_s$ is the strong coupling 
constant that depends on temperature~\cite{zantow}. 
%We have considered the effect of chemical potential to the production rate
%of strange quarks in the QGP phase. 

%%%%%%%%%%%%%%%%%%%%%%%%%%%%%%%%%%%%%%%%%%%%%%%%%%%%%%%%%%%%%%%%%%%%%
\begin{figure}
\begin{center}
\includegraphics[scale=0.45]{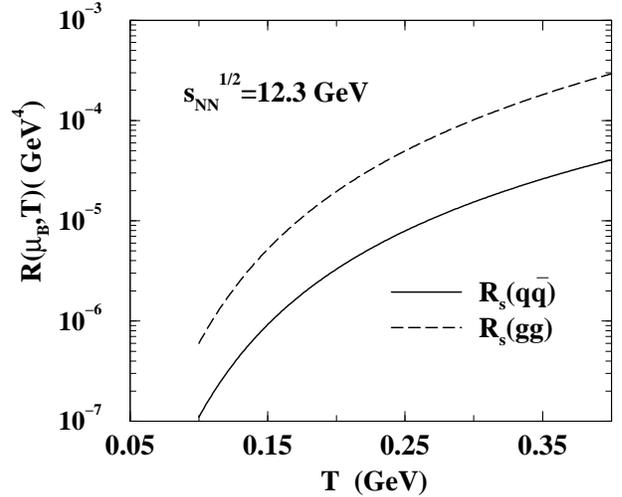}
\caption{Rate of production of $\bar{s}$ quark from $gg \rightarrow 
s \bar{s}$ and $q \bar{q} \rightarrow s \bar{s}$ with temperature.
} 
\label{fig1}
\end{center}
\end{figure} 
%%%%%%%%%%%%%%%%%%%%%%%%%%%%%%%%%%%%%%%%%%%%%%%%%%%%%%%%
\begin{figure}
\begin{center}
\includegraphics[scale=0.45]{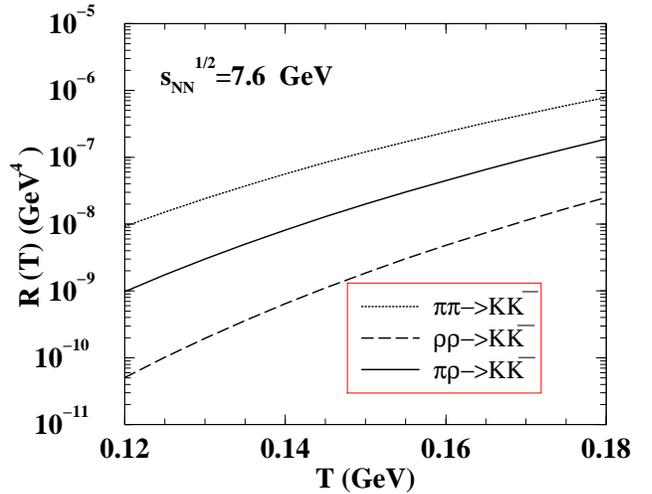}
\caption{The rate of kaon production from dominant meson meson   
interactions with temperature.}
\label{fig2}
\end{center}
\end{figure} 
%%%%%%%%%%%%%%%%%%%%%%%%%%%%%%%%%%%%%%%%%%%%%%%%%%%%%%%%
%%%%%%%%%%%%%%%%%%%%%%%%%%%%%%%%%%%%%%%%%%%%%%%%%%%%%%%
\subsection{$K^+$ and $K^-$ productions in the hadronic system}
%%%%%%%%%%%%%%%%%%%%%%%%%%%%%%%%%%%%%%%%%%%%%%%%%%%%%%%
The rate of $K^+(u\bar{s}$) and  $K^-(\bar{u}s$) productions in the 
hadronic phase can be categorized as due to (a) meson-meson ($MM$), 
(b) meson-baryon ($MB$) and 
(c) baryon- baryon ($BB$) interactions. 
In the present paper we quote only the main results for
kaon productions in the hadronic matter and 
refer to ~\cite{Brown} for details.

(a) For the first category $MM \rightarrow K\bar{K}$, we considered 
the following channels: 
$\pi \pi \rightarrow K \bar{K}$, $\rho \rho \rightarrow K \bar{K}$,  $\pi 
\rho \rightarrow K \bar{K^*}$ and $\pi \rho \rightarrow K^* \bar{K}$.
The invariant amplitude for these processes have been calculated 
from the following Lagrangians~\cite{Brown}. 
For the  $K^*K\pi$ vertex the interaction is given by,
\begin{equation}
{\cal L}_{K^*K\pi}=g_{K^*K\pi}K^{*\mu}\tau[K(\partial_{\mu}\pi)-
(\partial_{\mu}K)\pi]
\end{equation}
Similarly for  the $\rho K K$ vertex the interaction is, 
\begin{equation}
{\cal L}_{\rho KK}=g_{\rho KK}[K\tau(\partial_{\mu}K)-(\partial^
{\mu}K)\tau K]\rho^{\mu}
\end{equation}
The isospin averaged cross section ($\bar{\sigma}$) for $MM \rightarrow K \bar{K}$ (i.e., 
$\pi \pi \rightarrow K \bar{K}$, $\rho \rho \rightarrow K \bar{K}$ and $\pi 
\rho \rightarrow K \bar{K^*}$, $\pi \rho \rightarrow K^* \bar{K}$)
is evaluated by using, 
\begin{equation}
{\bar{\sigma}}=\frac{1}{32 \pi} \frac{P'}{sP} \int^1_{-1}dx M(s,x)
\end{equation}
where $P$ and $P'$ are
the three-momenta of the meson and kaons in the centre-of-mass frame, 
$x$ is the cosine of the angle between $P$ and $P'$. $M(s,x)$
is the iso-spin averaged squared invariant amplitude. 

(b) For meson baryon interactions the dominant channels are: $\pi\,N \rightarrow
\Lambda\,K$, $\rho\,N\rightarrow \Lambda\,K$, $\pi\,N\rightarrow N\,K\,\bar{K}$
and $\pi\,N\rightarrow N\,\pi\,K\,\bar{K}$. The isospin averaged 
cross section is given by~\cite{pdb}:

\begin{eqnarray}
\bar{\sigma}_{MB \rightarrow YK}=\sum_i\frac{(2J_i +1)}
{(2S_1+1)(2S_2+1)} \frac{4\pi}{k_i^2}\nonumber\\
\frac{\frac{\Gamma_i^2}{4}} 
{(s^{\frac{1}{2}}-m_i)^2+\Gamma_i^2/4} 
B_i^{in} B_i^{out}
\end{eqnarray}
$J_i$, $\Gamma_i$ and $m_i$ are 
the spin, width and mass of the resonances, 
($2S+1$) is the polarization states of the 
incident particles, $k$ is the centre of mass momentum of the initial state.
$B^{in}$ and $B^{out}$ are the branching ratios of initial and final 
state channels respectively. 
The index $i$ runs over all the resonance states. For interactions 
$\pi N \rightarrow \Lambda K$, $\rho N \rightarrow \Lambda K$ 
we have considered $N^*_1(1650), N^*_2(1710)$ and $N^*_3(1720)$ as the 
intermediate states. Values of various hadronic masses and
decay widths are taken from particle data book~\cite{pdb}. 

(c) For the last category of reactions {\it i.e.} for baryon baryon interactions
~\cite{wuko,randrup,brown2} 
the dominant processes  are: $N\,N\rightarrow N\,\Lambda\,K$,
$N\,\Delta\,\rightarrow N\,\Lambda\,K$,
$\Delta\,\Delta\,\rightarrow N\,\Lambda\,K$,
$N\,N\,\rightarrow N\,N\,K\,\bar{K}$,
$N\,N\,\rightarrow N\,N\,\pi\,\pi\,K\,\bar{K}$ and 
$N\,N\,\rightarrow N\,N\,\pi\,K\,\bar{K}$.

The isospin averaged cross section of kaon production 
from the process like $N_1 N_2 \rightarrow N_3 \Lambda K$  
is given by~\cite{wuko,randrup}   

\begin{eqnarray}
\bar{\sigma}_{N N \rightarrow N \Lambda K}&&=\frac{3 m_N^2}
{2 \pi^2 p^2 s} 
\int^{W_{max}}_{W_{min}} dW W^2 k 
\int^{q^{2}_+}_{q^2_-}dq^2  \nonumber \\
&&\frac{f^2_{\pi NN}}{m_{\pi}^2} F^2(q^2) \frac{q^2}{(q^2-m_{\pi}^2)^2}
{\bar \sigma_0 (W;q^2)}
\end{eqnarray}
pion is the intermediate particle for the above interaction, 
$m_N$ is the mass of $N$, $W$ is the total energy in the centre of 
mass system of pion and $N_2$, 
$W_{min}=m_K + m_\Lambda, W_{max}=s^{1/2}-m_N$ . 
$q_{\pm}^2=2 m_N^2-2EE' \pm 2pp'$ where $p$, $p^\prime$ 
are the momenta and $E$, $E^\prime$
are the  energies of $N_1$ and $N_3$ respectively. 
We take $f_{\pi NN}$ =1 and to constrain the finite size of the 
interaction vertices we use the form factor $F=
(\Lambda^2-m_\pi^2)/(\Lambda^2-q^2)$. 
$\bar{\sigma_0}$ is the isospin averaged cross section of $\pi N_2 \rightarrow 
\Lambda K$. Cross sections for the processes:  ${N \Delta \rightarrow 
N \Lambda K}$ 
and ${\Delta \Delta \rightarrow N \Lambda K}$ have been taken 
from ~\cite{randrup}. 
The cross section of other reactions {\it e.g.} 
$N\,N\,\rightarrow N\,N\,K\,\bar{K}$,
$N\,N\,\rightarrow N\,N\,\pi\,\pi\,K\,\bar{K}$ and 
$N\,N\,\rightarrow N\,N\,\pi\,K\,\bar{K}$
have been taken from  \cite{brown2}.
%Similarly the rate of productions for $K^-$ have been calculated considering 
%the same reactions as taken for $K^+$, apart from the interactions involving 
%hyperons.  
%following type of reactions:
%$MM \rightarrow K\bar{K}$, $MB \rightarrow BK\bar{K}$, 
%$MB \rightarrow MBK\bar{K}$, $BB \rightarrow BBK\bar{K}$, 
%$BB \rightarrow BBMK\bar{K}$, $BB \rightarrow BBMMK\bar{K}$,
%$KN \rightarrow KN \pi$, $KN \rightarrow KN \pi \pi$. 
In a baryon rich medium, $K^-$ gets absorbed due to its interaction
with the baryons. 
The reactions 
$K^- p \rightarrow \Lambda \pi^0$, $K^- p \rightarrow \sigma \pi^0$,
$K^- n \rightarrow \sigma p$, $K^- p \rightarrow \bar{K}^0 n$, 
$K^- n \rightarrow K^- n$
have been considered for $K^-$ absorption~\cite{brown2} in the nuclear
matter.
%The rates of kaon productions have been calculated by using Eq.~\ref{eq_rate} 
%with the above cross sections. 
%%%%%%%%%%%%%%%%%%%%%%%%%%%%%%%%%%%%%%%%%%%%%%%%%%%%%%%%
\begin{figure}
\begin{center}
\includegraphics[scale=0.45]{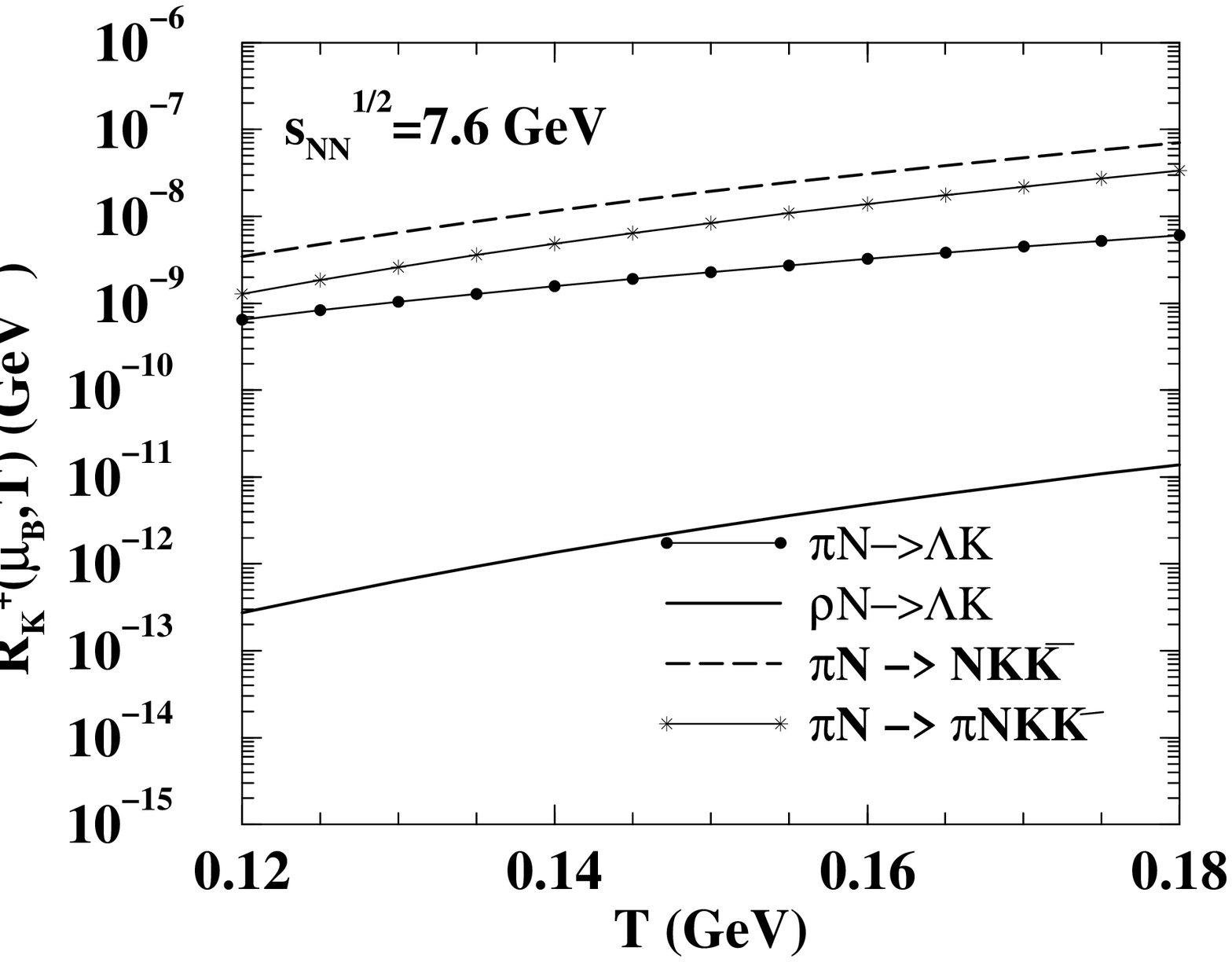}
\caption{Rate of kaon productions from the meson-baryon interactions 
 with temperature.
}
\label{fig3}
\end{center}
\end{figure} 
%%%%%%%%%%%%%%%%%%%%%%%%%%%%%%%%%%%%%%%%%%%%%%%%%%%%%%%%
\begin{figure}
\begin{center}
\includegraphics[scale=0.45]{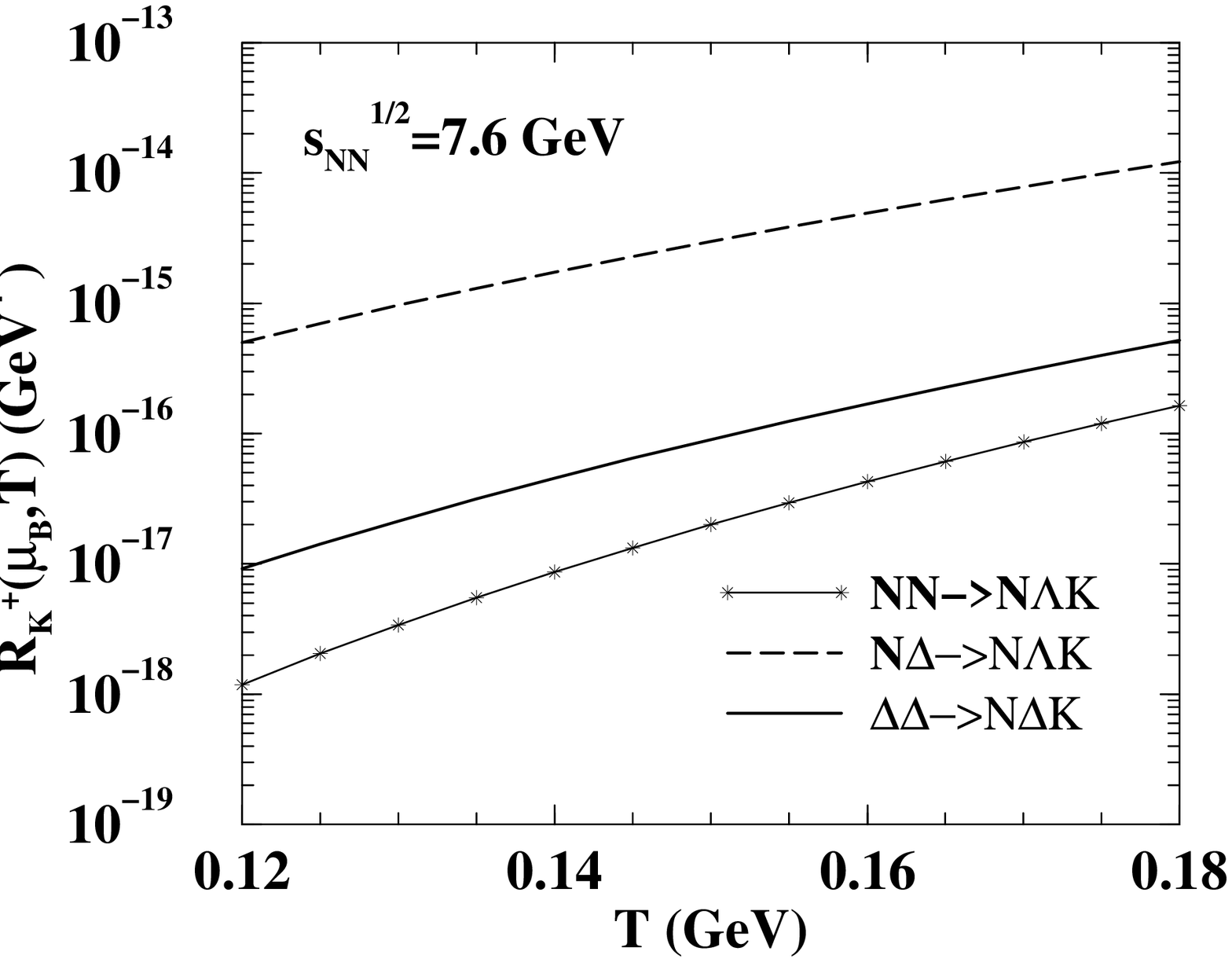}
\caption{Rate of kaon productions from baryon baryon  interactions 
with temperature.
}
\label{fig4}
\end{center}
\end{figure} 
%%%%%%%%%%%%%%%%%%%%%%%%%%%%%%%%%%%%%%%%%%%%%%%%%%%%%%%%
\begin{figure}
\begin{center}
\includegraphics[scale=0.45]{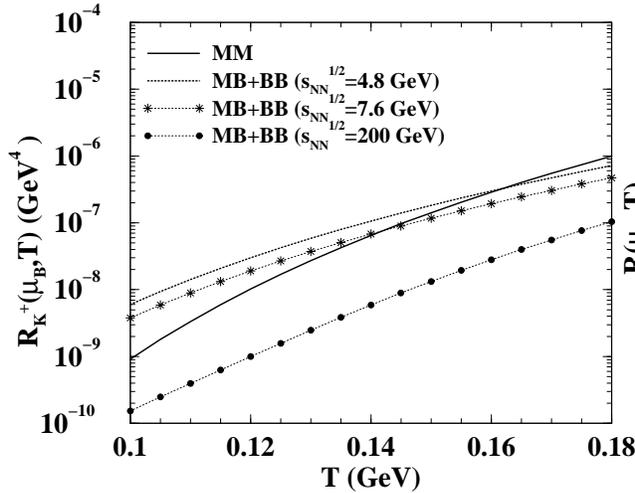}
\caption{Rate of Kaon productions from meson-meson(MM) interactions, 
 meson-baryon(MB) and baryon-baryon(BB) interactions at 
different center of mass energies}
\label{fig5}
\end{center}
\end{figure} 
%%%%%%%%%%%%%%%%%%%%%%%%%%%%%%%%%%%%%%%%%%%%%%%%%%%%%%%%
\begin{figure}
\begin{center}
\includegraphics[scale=0.45]{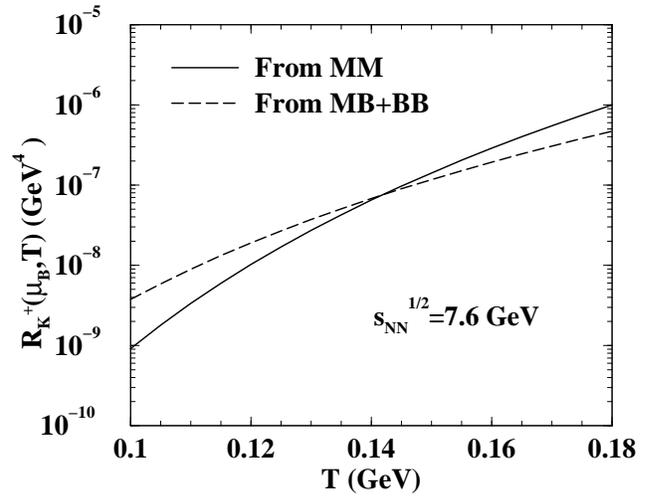}
\caption{Comparison between rates of kaon productions from MM and
MM +  MB interactions with temperature.
}
\label{fig6}
\end{center}
\end{figure} 
%%%%%%%%%%%%%%%%%%%%%%%%%%%%%%%%%%%%%%%%%%%%%%%%%%%%%%%%
\begin{figure}
\begin{center}
\includegraphics[scale=0.45]{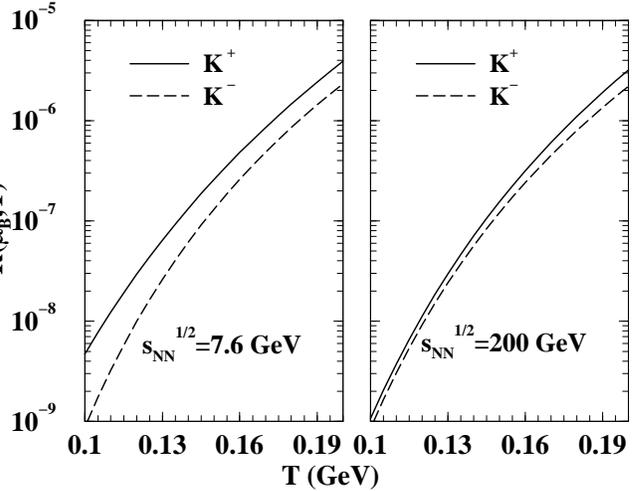}
\caption{Total $K^+$ and $K^-$ production rates with temperature at 
center of mass energy=7.6 GeV and 200 GeV.} 
\label{fig7}
\end{center}
\end{figure} 
%%%%%%%%%%%%%%%%%%%%%%%%%%%%%%%%%%%%%%%%%%%%%%%%%%%%%%%
\subsection{Rate of strangeness productions}
%%%%%%%%%%%%%%%%%%%%%%%%%%%%%%%%%%%%%%%%%%%%%%%%%%%%%%%
$dN/d^4x(\equiv R)$, 
the number of $s$ quarks produced per unit time per unit volume 
at temperature $T$ and baryonic chemical potential $\mu_B$ is given by 
\begin{equation}
\frac{dN}{d^4x}=\int \frac{d^3p_1}{(2 \pi)^3}f(p_1) 
\int  \frac{d^3p_2}{(2 \pi)^3}f(p_2) v_{rel} \sigma
\label{eq_rate}
\end{equation}
where $p_i$'s are the momenta of the incoming particles and $f(p_i)$'s are
the respective phase space distribution functions (through which the 
dependence on $T$ and $\mu_B$ are introduced), 
$v_{\mathrm rel} = |{v_1-v_2}|$ is the 
relative velocity  of the incoming particles and $\sigma$ is the
production cross sections for the reactions. 
The same equation can be used for kaon production by appropriate
replacements of phase space factor and cross sections.

%%%%%%%%%%%%%%%%%%%%%%%%%%%%%%%%%%%%%%%%%%%%%%%%%%%%%%%%%%%%%%%%%%%%%%%%
\section{Evolution of Strangeness}
%%%%%%%%%%%%%%%%%%%%%%%%%%%%%%%%%%%%%%%%%%%%%%%%%%%%%%%%%%%%%%%%%%%%%%%%
The possibility of formation of 
a fully equilibrated system  in high energy nuclear 
collisions is still a fiercely debated
issue because of the finite size and life time of  system.
%However some experimental 
%data at higher centre of mass energies supports thermalization of the
%system  where the high density of the produced 
%particles may overshoot the effect of finite-size. 
In the present work we assume that the strange quarks or 
the strange hadrons (depending on the value of 
$\sqrt{s_{\mathrm {NN}}}$) produced as a result of the collisions
are not in chemical equilibrium.
The time evolution of the strangeness in either QGP or hadronic
phase is governed by the momentum integrated Boltzmann equation.
We have  assumed that the initial density
of strange quarks or kaons (depending on the initial
conditions (I) or (II)) is  $20\%$ away from the corresponding 
equilibrium density. We will comment on the amount of deviations
from chemical equilibrium later. 

\subsection{Evolution in QGP and hadronic phase}
The momentum integrated Boltzmann equation has been applied to study
the freeze-out of elementary particles during the thermal expansion of the
early universe~\cite{kolbandturner}. In the present work we follow
similar procedure to study 
the evolution of the strange quarks and anti-quarks in the QGP phase or kaons
in the hadronic phase.  The coupled equations describing
the evolution of $i$ (particle) 
and $j$ (anti particle) with proper time $\tau$ is given by:
\begin{eqnarray}
\frac{dn_i}{d\tau} &=& R_{i}(\mu_B, T)
[ 1 - \frac{n_i n_j}{n_{i}^{eq}{n_{j}^{eq}}}] - \frac{n_{i}}
{\tau} \nonumber \\
\frac{dn_{j}}{d\tau} &=& R_{j}(\mu_B, T)
[ 1 - \frac{n_j n_{i}}{n_{j}^{eq}{n_{i}^{eq}}}] - \frac{n_{j}}
{\tau} .
\label{eqn_evol1}
\end{eqnarray}
where, $n_i$ ($n_j$) and ${n_i}^{eq}$ (${n_j}^{eq}$) are the non-equilibrium 
and equilibrium densities of $i$ ($j$) type of particles respectively.
$R_i$ is the rate of production of particle $i$ 
at temperature $T$ and chemical potential 
$\mu_B$, $\tau$ is the proper time. 
First term on the right hand side of Eq. \ref {eqn_evol1} is the production 
term and the second term represents  the dilution of the system 
due to expansion. The variation of temperature and the baryonic chemical
potential with time is governed by the hydrodynamic equations
(next section).
The indices $i$ and $j$ in Eq.\ref{eqn_evol1} are replaced  
by $s, \bar s$ quark 
in the QGP phase  and by $K^+, K^-$ in the hadron phase respectively. 

\subsection{Evolution in the mixed phase}
For higher colliding energies 
i.e., $\sqrt{s}$ $\ge$ 8.76 GeV an initial partonic phase is assumed.
The hadrons are formed at a transition temperature, $T_c=190$ MeV 
through a first order phase transition from QGP to hadrons.
The fraction of the QGP phase in the mixed phase at a proper time $\tau$
is given by~\cite{kapusta,japr}:
\begin{equation}
f_Q(\tau)=\frac{1}{r-1}(r\frac{\tau_H}{\tau}-1)
\end{equation}
where
$\tau_Q$ ($\tau_H$) is the time at which the QGP 
(mixed) phase ends, $r$ is the ratio of statistical degeneracy 
in QGP to hadronic phase.  The evolution of the kaons are governed by
~\cite{kapusta}:
\begin{eqnarray}
%\frac{dn_{\bar{s}}}{d\tau} &=& R_{\bar{s}}(\mu, T_c)
%[ 1 - \frac{n_s n_{\bar{s}}}{n_{s}^{eq}{n_{\bar{s}}^{eq}}}] - \frac{n_{\bar{s}}}
%{\tau} \nonumber \\
%\frac{dn_{s}}{d\tau} &=& R_{s}(\mu, T_c)
%[ 1 - \frac{n_s n_{\bar{s}}}{n_{s}^{eq}{n_{\bar{s}}^{eq}}}] - \frac{n_{s}}
%{\tau} \nonumber \\
\frac{dn_{K^+}}{d\tau}& = &R_{K^+}(\mu_B,T_c)
[ 1 - \frac{n_{K^+}n_{K^-}}{n_{K^+}^{eq}n_{K^-}^{eq}}] - \frac{n_{K^+}}
{\tau}+ \nonumber \\
&&\frac {1}{f_H} \frac {d f_H}{d \tau} \left(\delta n_{\bar{s}}-n_{K^+}\right) 
\nonumber\\
\frac{dn_{K^-}}{d\tau}& = &R_{K^-}(\mu_B,T_c)
[ 1 - \frac{n_{K^+}n_{K^-}}{n_{K^+}^{eq}n_{K^-}^{eq}}] - \frac{n_{K^-}}
{\tau} +\nonumber \\
&&\frac {1}{f_H} \frac {df_H}{d \tau} \left(\delta n_{s}-n_{K^-}\right) 
\label{eqmix}
\end{eqnarray}
Similar equation exist for the evolution of $s$ and $\bar{s}$ quarks in the
mixed phase (see~\cite{kapusta} for details).
In the above equations $f_H(\tau)=1-f_Q(\tau)$ 
represents  the fraction of hadrons in 
the mixed 
phase at time $\tau$. The last term stands for the hadronization of 
$\bar{s} (s)$ quarks to $K^+ ( K^- )$~\cite{kapusta,matsui}. Here $\delta$ 
is a parameter which indicates the fraction of $\bar{s}(s)$  quarks 
hadronizing to $K^+(K^-)$. 
$\delta=0.5$ indicates the formation
of $K^+$ and $K^0$  in the mixed phase because half of the $\bar{s}$
form $K^+$ and the rest hadronize to $K^0$.
%The value of $\delta=0.5$ if we consider  
%only $K^+$ and $K^0$ formation in the mixed phase.  
%The analogous evolution of $s$ and $\bar{s}$ can be written down (see
%Ref.~\cite{kapusta} for details).

%For equations similar to Eq.~\ref{eqmix} describing the  evolution of the
%strange quarks and antiquarks  in the mixed phase 
%see Ref~\cite{kapusta}.
%%%%%%%%%%%%%%%%%%%%%%%%%%%%%%%%%%%%%%%%%%%%%%%%%%%%%%%%
%\begin{figure}
%\begin{center}
%\includegraphics[scale=0.45]{k+_energy.eps}
%\caption{kaon and pion production with centre of mass energy.} 
%\label{fig5}
%\end{center}
%\end{figure} 
%%%%%%%%%%%%%%%%%%%%%%%%%%%%%%%%%%%%%%%%%%%%%%%%%%%%%%%%%%%%%%%%%%%%%%%%%
\subsection{Space time evolution}
%%%%%%%%%%%%%%%%%%%%%%%%%%%%%%%%%%%%%%%%%%%%%%%%%%%%%%%%%%%%%%%%%%%%%%%%%
The partonic/hadronic system produced in nuclear 
collisions evolves in space-time. The space-time evolution of the 
bulk matter is governed by the relativistic hydrodynamic equation:
\be
\partial_\mu T^{\mu\nu}=0
\label{hydro1}
\ee
with boost invariance along the longitudinal direction~\cite{jdb}.
In the above equation $T^{\mu\nu}=(\epsilon+P)u^\mu u^\nu-g^{\mu\nu}P$,
is the energy momentum tensor for ideal fluid,
$\epsilon$ is the energy density, $P$ is the pressure
and $u^\mu$ is the hydrodynamic
four velocity.
The net baryon number conservation in the system
is governed by:
\be
\partial_\mu (n_Bu^\mu)=0
\label{hydro2}
\ee
where $n_B$ is the net baryon density.
Eqs.~\ref{hydro1} and~\ref{hydro2} have been solved (see ~\cite{biro,ijmpa}
for details) to obtain the variation of temperature and baryon density
with proper time. The initial temperatures corresponding to different
$\sqrt{s_{\mathrm {NN}}}$ are taken from Table I. The baryonic chemical 
potential at freeze-out are taken from the parametrization of $\mu_B$ with 
$\sqrt{s_{\mathrm NN}}$ ~\cite{ristea}(see also~\cite{andronic}) and the 
baryonic chemical potential at the initial state is obtained from the net 
baryon number conservation equation.

%%%%%%%%%%%%%%%%%%%%%%%%%%%%%%%%%%%%%%%%%%%%%%%%%%%%%%%%%%%%%%%%%%%%%%%%%%

The initial temperatures of the systems formed after
nuclear collisions have been evaluated from the
measured hadronic multiplicity, $dN/dy$ by using the
following relation: 
\begin{equation}
T_i^3=\frac{2 \pi^4}{45 \zeta (3)}
         \frac{1}{\pi R^2 \tau _i}
         \frac{90}{4\pi^2g_{eff}}\frac{dN}{dy},
\end{equation}
where $\zeta(3)$ denotes the Riemann zeta function, $R$ is the transverse 
radius [$\sim$ $1.1(N_{part}/2)^{1/3}$, $N_{part}$ is the number of 
participant nucleons ] of the colliding system , ${\tau}_i$ is the initial 
time and $g_{eff}$ is the statistical degeneracy. Initial 
temperatures for different $\sqrt{s_{NN}}$ are tabulated in table I. 
%%%%%%%%%%%%%%%%%%%%%%%%%%%%%%%%%%%%%%%%%%%%%%%%%%%%%%%%%%%%%%%%%%%%%%%%%
\begin{table}[h]
\caption{Initial conditions for the transport calculation. Colliding energies 
are in centre of mass frame} 
\begin{tabular}{lccccr}
\tableline 
$\sqrt{s_{\mathrm {NN}}}$ (GeV)&$T_i$ (GeV)& $T_c$ (GeV)\\
\tableline
3.32& 0.115& -\\
3.83& 0.128& -\\
4.8& 0.150& -\\
6.27& 0.160& -\\
7.6& 0.187& -\\
8.76& 0.210& 0.190\\
12.3& 0.225& 0.190\\
17.3& 0.25& 0.190\\
62.4& 0.3& 0.190\\
130& 0.35& 0.190\\
200& 0.40& 0.190\\
%5500& 0.650& 0.190\\
\tableline
\end{tabular}
\label{tableII}
\end{table}
%%%%%%%%%%%%%%%%%%%%%%%%%%%%%%%%%%%%%%%%%%%%%%%%%%%%%%%%%%%%%%%%%%%%%%%%%%%
\section{Results and Discussion}
%%%%%%%%%%%%%%%%%%%%%%%%%%%%%%%%%%%%%%%%%%%%%%%%%%%%%%%%%%%%%%%%%%%%%%%%%
The variation of the 
number of strange anti-quarks produced per unit volume 
per unit time  with temperature 
has been displayed in Fig.~\ref{fig1} for a baryonic chemical potential
$\mu_q=107$ MeV. 
It is observed that the process of gluon fusion dominates over the 
$q\bar{q}$ annihilation for the entire temperature range under consideration,
primarily because at high $\mu_B (=3\mu_q)$ the number of anti-quarks is suppressed.
In Fig.~\ref{fig2}, the production rate of $K^+$ from the
$MM\rightarrow K\bar{K}$ type of reactions has been depicted
for $\sqrt{s_{\mathrm {NN}}}=7.6$ GeV.
The production rate from pion annihilation dominates over 
the reactions that involves $\rho$ mesons, because 
the thermal phase space factor of $\rho$ is small due its
larger mass compared to pions and smaller production 
cross section. Results for interactions 
involving mesons and baryons are displayed in Fig.~\ref{fig3}. 
It is observed that the interactions involving pions and
nucleons in the initial channels dominate over that 
which has a $\rho$ meson in the incident channel.
In fact, contributions from the reactions $\rho\,N\rightarrow \Lambda\,K$ 
has negligible effect on the total productions from the 
meson baryon interactions.  
The kaon production from baryon-baryon interaction is displayed
in Fig.~\ref{fig4}.
The contributions from $N\,\Delta\rightarrow N\,\Lambda\,K$ dominates
over the contributions from 
$N\,N\rightarrow N\,\Lambda\,K$  and
$\Delta\,\Delta\rightarrow N\,\Delta\,K$ for the 
temperature range $T=120$ to 180 MeV.  
%It has been investigated earlier in \cite{randrup, aichelin} 
%that $BB \rightarrow BYK$ has a dominant contribution to the rate at 
%Bevelac energies but here at these energies (AGS-RHIC) the contribution 
%from $BB \rightarrow BYK$  is not that important. 

In Fig.\ref{fig5} the rates of $K^+$ productions from 
meson-meson interactions has been compared with those 
involving baryons {\it i.e.} with meson-baryon 
and baryon-baryon interactions for different $\sqrt{s_{\mathrm {NN}}}$ 
(different $\mu_B$).
The results clearly indicate the dominant role of baryons at
lower collision energies which diminishes with increasing 
$\sqrt{s_{\mathrm {NN}}}$. At low temperature the baryonic contribution
is more than the mesonic one for lower beam energy. Rate of productions 
(from MB+BB interactions) at $\sqrt{s_{\mathrm {NN}}}$=4.8 GeV is more 
compared to the rates at $\sqrt{s_{\mathrm {NN}}}$=7.6 and 200 GeV, since 
$\mu_B$ at $\sqrt{s_{\mathrm {NN}}}$=4.8 GeV is more (see table-II). 
Production rate from pure mesonic interactions does not depend on $\mu_B$ 
hence same for all. It is quite clear from the 
results displayed in Fig.\ref{fig5} that more the 
baryonic chemical potential (lower the centre of mass energy), more is 
the rate from BB and MB interactions compared to MM interactions. For 
a system having lower chemical potential (higher centre of mass energy) 
the rate of production from mesonic interactions is dominant.
%%%%%%%%%%%%%%%%%%%%%%%%%%%%%%%%%%%%%%%%%%%%%%%%%%%%%%%%%%%%%%%%%%%%%%%%%%%
\begin{table}[h]
\caption{ Chemical potential for different centre of mass energies}
\begin{tabular}{cc}
\tableline
Center of mass energy$(\sqrt s_{NN})$& Chem. potential ($\mu_{B}$)\\
 (in  A GeV) & (MeV)\\
\tableline
3.32&595  \\
3.83& 568  \\
4.8& 542  \\
6.27&478 \\
7.6& 432\\
8.76& 398\\
12.3& 321\\
17.3& 253\\
62.4&86\\
130& 43\\
200& 28\\
\tableline
\end{tabular}
\label{tableI}
\end{table}
%%%%%%%%%%%%%%%%%%%%%%%%%%%%%%%%%%%%%%%%%%%%%%%%%%%%%%%%%%%%%%%%%%%%%%%%%%%
A comparison is made between rates of kaon productions from meson meson 
interactions (MM) and meson-baryon (MB) plus baryon-baryon (BB) interactions 
for $\sqrt{s_{\mathrm {NN}}}=7.6$ GeV. At this energy baryons  
and mesons are equally important as shown in Fig.~\ref{fig6}. 

In Fig.~\ref{fig7} the net rates of productions for $K^+$ and
$K^-$ have been depicted for $\sqrt{s_{\mathrm {NN}}}= 7.6$ GeV (left panel)
and 200 GeV (right panel). At $\sqrt{s_{\mathrm {NN}}}=7.6$ GeV the production 
of $K^+$ dominates over $K^-$ for the entire temperature range.  However, 
for large $\sqrt{s_{\mathrm {NN}}}$ (low $\mu_B$)
the productions of $K^+$ and $K^-$ are similar. 
%At lower energies the system has higher baryonic chemical potential 
%with more net baryons(nucleons). Hence more $K^-$ get absorbed leading 
%to less productions compared to $K^+$. But at higher energies $\mu_B$ is 
%less, so less $K^-$ get absorbed resulting a similar production as $K^+$. 
The strong absorption of the $K^-$ by nucleons in a baryon
rich medium resulting in lower production yield of $K^-$ compared
to $K^+$.
This may be contrasted  with the experimental findings of BRAHMS experiment
~\cite{BRAHMS} where it is observed that at mid-rapidity (small $\mu_B$ 
due to nuclear transparency at RHIC energy) the $K^+$ and $K^-$ yields are
similar but at large rapidity (large $\mu_B$) $K^-$ yield is smaller
than $K^+$ due to large $K^-$ nucleon absorption. 

\begin{figure}
\begin{center}
\includegraphics[scale=0.45]{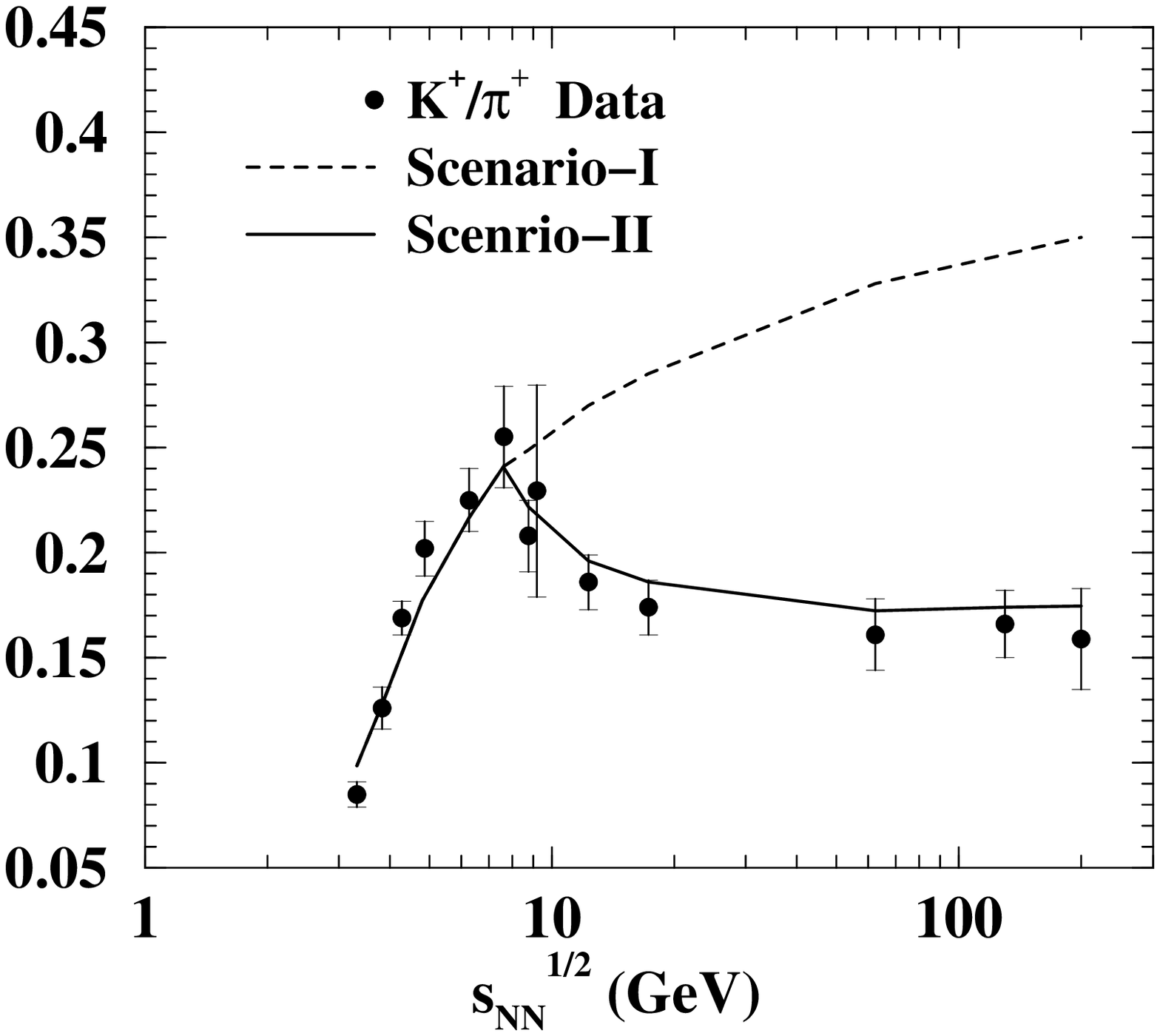}
\caption{$K^+$/$\pi^+$ ratio for different centre of mass energies.
Scenario - I 
represents for pure initial hadronic scenario for all centre of mass 
energies. Scenario - II represents for the calculation with hadronic initial 
conditions for low $\sqrt{s_{NN}}$ and partonic initial conditions for 
higher $\sqrt{s_{NN}}$. See text for details.} 
\label{fig8}
\end{center}
\end{figure}

In Fig.~\ref{fig8} the variations of $R^+$ 
with $\sqrt{s_{\mathrm {NN}}}$ are depicted. The experimental data on $R^+$ 
is well reproduced if a partonic initial phase (scenario-II) is assumed beyond
$\sqrt{s_{\mathrm {NN}}}$=8.7 GeV. A ``mindless'' extrapolation of hadronic
initial state (scenario-I)for all the $\sqrt{s_{\mathrm {NN}}}$ up to RHIC
energy show an  increasing trend in disagreement with
the experimental data at higher $\sqrt{s_{\mathrm {NN}}}$. 
In both the scenarios, I and II, the curves at higher 
$\sqrt{s_{\mathrm {NN}}}$ (RHIC energies) becomes flatter. 
That is because at higher energies the $K^+$ productions in the 
hadronic phase are dominated by mesonic interactions and the production rates 
from mesons are same for all $\sqrt{s_{\mathrm {NN}}}$ for a given temperature 
range. But at lower energies the rates of kaon productions are dominated by the 
effective interactions among the baryonic degrees of freedom. 
The composition of matter formed in heavy ion collision changes from a 
matter dominated by baryons to a matter dominated by mesons with the increase  
in colliding energy. The $\mu_B$ changes from 86 MeV to 28 MeV as 
$\sqrt{s_{\mathrm {NN}}}$ varies from 62.4 GeV to 200 GeV 
(Table II). The change in the $K^+$ production in the hadronic phase due to the 
change in $\mu_B$ mentioned above is marginal - resulting in the flatness in
$R^+$ at higher energies. The decrease of the value of 
the $R^+$  beyond $\sqrt{s_{\mathrm {NN}}}$=7.6 GeV showing 'horn' like 
structure happens only when an initial partonic phase is considered.  
Such a non-monotonic behaviour of $R^+$
can be understood as due to larger entropy productions from 
the release of large colour degrees of freedom (resulting
in more pions yield) compared to strangeness beyond energy 7.6 GeV .

\begin{figure}
\begin{center}
\includegraphics[scale=0.45]{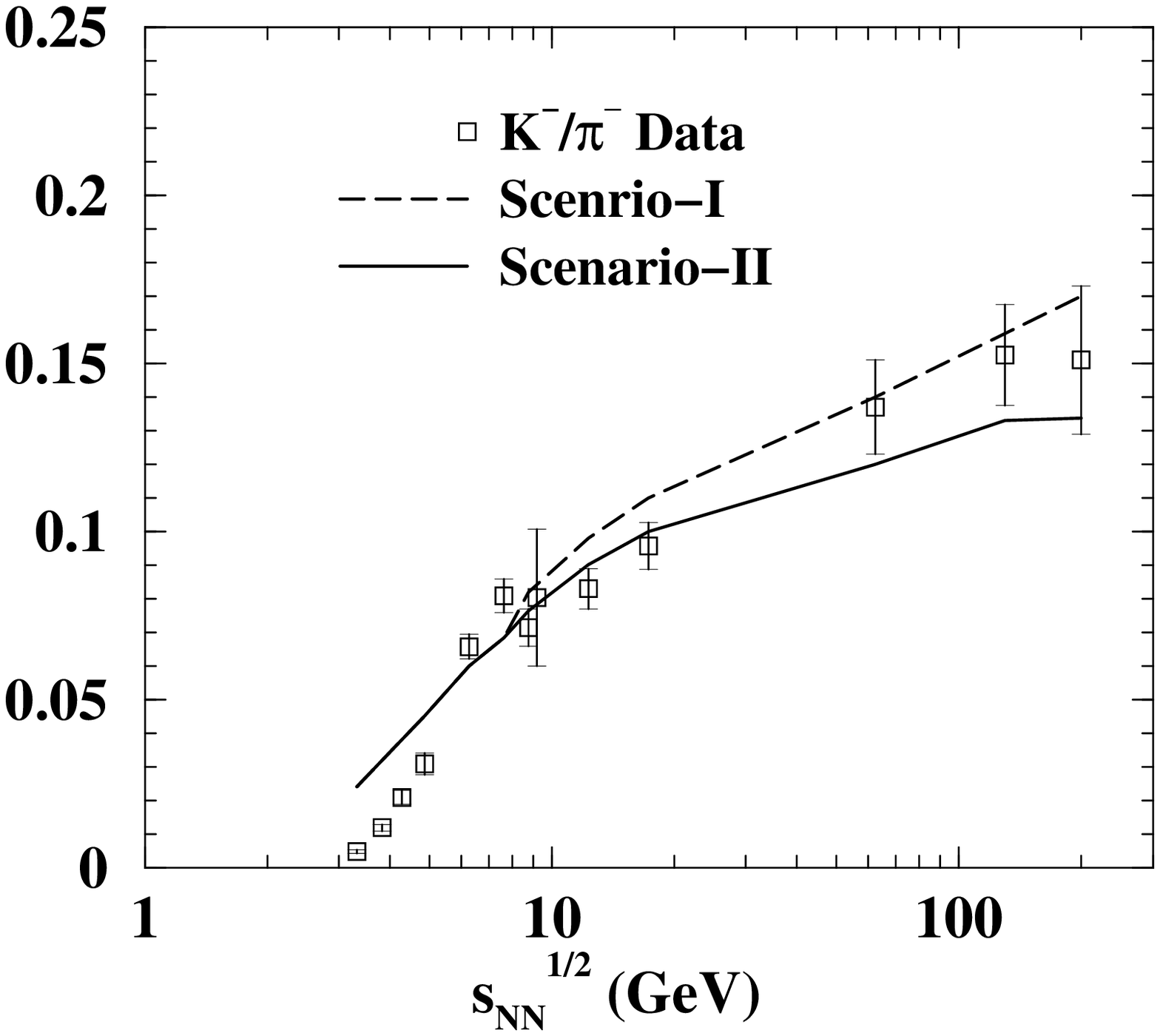}
\caption{$K^-$/$\pi^-$ ratio for different centre of mass energies. Scenario-I 
represents for pure initial hadronic scenario for all centre of mass 
energies. Scenario-II represents for the calculation with hadronic initial 
conditions for low $\sqrt{s_{NN}}$ and partonic initial conditions for 
higher $\sqrt{s_{NN}}$. See text for details.} 
\label{fig9}
\end{center}
\end{figure}

In Fig.~\ref{fig9}, the variations of $R^-$ with 
$\sqrt{s_{\mathrm {NN}}}$ is  displayed. 
$R^-$ has a lower value 
compared to $R^+$ at lower energies  
since $K^-$ get absorbed in the baryonic medium. At higher energies  
$K^-$ is closer to
$K^+$ because production of $K^+$ and $K^-$ is similar 
in baryon free medium, which may be realized at higher 
collision energies.

%%%%%%%%%%%%%%%%%%%%%%%%%%%%%%%%%%%%%%%%%%%%%%%%%%%%%%%%%%%%%%%%%%%%%
%%%%%%%%%%%%%%%%%%%%%%%%%%%%%%%%%%%%%%%%%%%%%%%%%%%%%%%%%%%%%%%%%%%%%%
%%%%%%%%%%%%%%%%%%%%%%%%%%%%%%%%%%%%%%%%%%%%%%%%%%%%%%%%%%%%%%%%%%%%%%
%%%%%%%%%%%%%%%%%%%%%%%%%%%%%%%%%%%%%%%%%%%%%%%%%%%%%%%%%%%%%%%%%%%%%%
\begin{figure}
\begin{center}
\includegraphics[scale=0.45]{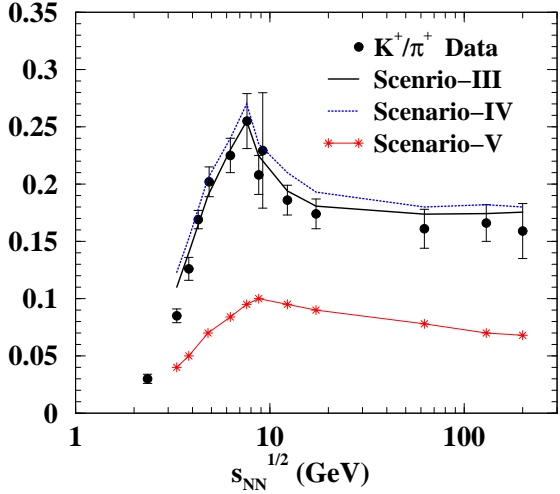}
\caption{$K^+$/$\pi^+$ ratio for different centre of mass energies. 
Scenario-III assumes complete equilibrium of strange quarks and hadrons. 
The production through secondary processes have been ignored.
Scenario IV is same as III with secondary productions processes are on and
scenario V represents zero strangeness initially but secondary productions
are switched on.
} 
\label{fig10}
\end{center}
\end{figure} 
%%%%%%%%%%%%%%%%%%%%%%%%%%%%%%%%%%%%%%%%%%%%%%%%%%%%%%%%%%%%%%%%%%%%%%
In Fig.~\ref{fig10} the $R^+$ is depicted as a function of 
$\sqrt{s_{\mathrm NN}}$ for other scenarios (III, IV and V).  
on the one hand 
when the strange quarks and kaons are formed
in complete equilibrium but their secondary productions are neglected 
during the evolution (scenario III) then the data is well reproduced.
On the other hand, in the scenario (IV) 
when the system is formed in equilibrium (as in III) but the 
productions of strange quarks and kaons are switched on 
through secondary processes then
the data is slightly overestimated at high $\sqrt{s_{\mathrm NN}}$.
However, we have seen that the data is also reproduced well
in the scenario II as discussed above. 
This indicate that the deficiency of strangeness below    
its equilibrium value as considered in (II) is 
compensated by the secondary productions. 
In scenario V we assume that vanishing initial strangeness
and observed that the production of strangeness throughout
the evolution is not sufficient to reproduce the data. The productions
from secondary processes are small but not entirely negligible (V).
%%%%%%%%%%%%%%%%%%%%%%%%%%%%%%%%%%%%%%%%%%%%%%%%%%%%%%%%%%%%%%%%%%%%%%
\begin{figure}
\begin{center}
\includegraphics[scale=0.45]{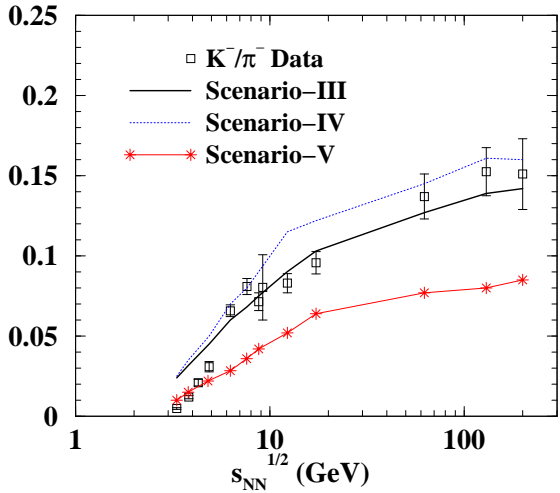}
\caption{Same as \protect{Fig.~\ref{fig10}} for $K^-$/$\pi^-$. 
} 
\label{fig11}
\end{center}
\end{figure} 
%%%%%%%%%%%%%%%%%%%%%%%%%%%%%%%%%%%%%%%%%%%%%%%%%%%%%%%%%%%%%%%%%%%%%%
In Fig.~\ref{fig11} the $R^-$ has been displayed as a function of 
$\sqrt{s_{\mathrm NN}}$.  A trend similar to the results shown in 
Fig.~\ref{fig10} is observed. The data is overestimated 
for the intermediate $\sqrt{s_{\mathrm NN}}$ in the scenario IV,
reproduced well in scenario III and underestimated for
the scenario V.

\section{Summary and Conclusions}
The evolution of the strangeness in the system formed in 
nuclear collisions at relativistic energies have been
studied within the framework of momentum integrated Boltzmann
equation. The Boltzmann equation has been used to study 
the evolution of $s$ and $\bar{s}$ in the partonic phase and  $K^-$ 
and $K^+$ in the  hadronic 
phase. The calculation has been done for different centre of mass 
energies ranging from AGS to RHIC. We get a non-monotonic variation 
of $K^+/\pi^+$ with $\sqrt{s_{NN}}$  when an 
initial partonic phase is assumed  for $\sqrt{s_{NN}}=8.76$ GeV and beyond. 
A monotonic rise of $K^+/\pi^+$ is observed when a pure hadronic scenario 
is assumed for all centre of mass energies. The $K^-/\pi^-$ data is 
unable to differentiate between the two initial conditions mentioned before.

%We argue that the experimentally observed horn of
%the $K^+/\pi^+$ with colliding energy appears due to the release of large
%number of colour degrees of freedom which contributes to the net entropy
%content of the system beyond certain $\sqrt{S_{NN}}$.  

Some comments on the values of the initial parameter are
in order at this point. We have seen that a $10\%$ variation
in the initial temperature does not change the results  
drastically.  
We have  assumed that the initial density
of strange quarks or kaons depending on the scenario (I) or (II)
is about $20\%$ away from the corresponding equilibrium density.
Results from  a scenario where  strange quarks or  kaons are formed in
complete equilibrium  and the production is ignored during the evolution
then the data is well reproduced (scenario III). 
If the the strangeness is produced in equilibrium 
and the production is included during the expansion
stage then the data is overestimated.
However, if the system is formed with zero strangeness then the  
theoretical results underestimate the data substantially. 
This indicate that the production of strangeness during the
expansion of the system is small but not entirely negligible.
The deficiency assumed in scenario II is compensated by the
production during evolution.

\par 
{\bf Acknowledgment:} JA and SB are  supported by DAE-BRNS
project Sanction No.  2005/21/5-BRNS/2455. Thanks for B. Mohanty and Lokesh 
Kumar for providing the experimental data.

%%%%%%%%%%%%%%%%%%%%%%%%%%%%%%%%%%%%%%%%%%%%%%%%%%%%%%%%%%%%%%%%%%%%%%

\end{document}